\documentstyle[sprocl]{article}

\include{psfig}

\input epsf

\begin{document}

\title{
WHY $N^*$'s ARE IMPORTANT
}

\author{NATHAN ISGUR}
\address{Jefferson Lab, 12000 Jefferson Avenue, Newport News, U. S. A.}

% You may repeat \author and \address if necessary

\maketitle

\abstracts{
The study of $N^*$'s can provide us with critical insights
into the nature of QCD in the confinement domain. The keys to progress
in this domain are the identification of its important  degrees of 
freedom and the effective forces between them.  I report on the growing evidence in
support of the flux tube model, and comment on the connection between this model and
spontaneous chiral symmetry breaking, on the spin-dependence of the long-range confining
potential, on the evidence for short-range one gluon exchange, on instantons, and on the
one pion exchange model. }

%%%%%%%%%%%%%%%%%%%%%%%%%%%%%%%%%%%%%%%%%%%%%%%%%%%%%%%%%%%%%%%%%%
\section{Why $N^*$'s?}
%%%%%%%%%%%%%%%%%%%%%%%%%%%%%%%%%%%%%%%%%%%%%%%%%%%%%%%%%%%%%%%%%%

    There are three main reasons (one for each quark) why I
believe that $N^*$'s
deserve the special attention they get from the series of 
Workshops to
which $N^*2000$ belongs:

\medskip

\noindent $\bullet$  The first is that nucleons {\it are}
the stuff of which our world is made. As such they must be at  the center
of any discussion  of why the world we
actually experience has the character it does. I am convinced that completing this
chapter in the history  of science will be one of the most interesting and fruitful
areas of physics for at least the next thirty years.

\noindent $\bullet$    My second reason is that they are the simplest system in
which the quintessentially nonabelian character of QCD is manifest. There
are, after all,  $N_c$ quarks in a proton because there are
$N_c$ colors.

\noindent $\bullet$      The third reason is that history has taught us that, while
relatively simple, baryons are sufficiently complex to reveal physics hidden from us in
the mesons. There are many examples of this, but  one famous example
should suffice: Gell-Mann and Zweig were forced to the quarks by 
$3\times 3\times 3$ giving the octet and decuplet, while mesons admitted of many possible
solutions.

%%%%%%%%%%%%%%%%%%%%%%%%%%%%%%%%%%%%%%%%%%%%%%%%%%%%%%%%%%%%%%%%%%
\section{What are the Key Issues?}
%%%%%%%%%%%%%%%%%%%%%%%%%%%%%%%%%%%%%%%%%%%%%%%%%%%%%%%%%%%%%%%%%%

    I am convinced that the keys to a 
qualitative understanding of ``strong QCD"~\cite{Close} 
are the same as in most other areas of physics: identifying the
appropriate degrees of freedom and the 
effective forces between them.

    Let me  remind you why the basic degrees of freedom of
QCD and the elementary gluon-mediated interactions between them are {\it
not} useful starting points for understanding QCD in the confinement
regime. It is not difficult to demonstrate that the string tension $b$ of QCD must depend
on the strong coupling constant $g$ according to
\begin{equation}
b \propto  exp \left[ - {16\pi^2\over {11g^2}}\right]~~.
\end{equation}
Almost all other interesting properties (including the $N^*$ masses) have this same
$e^{-1/g^2}$ dependence. This essential singularity in $g^2$ means that the ``Feynman
diagrammar" is useless: {\it plane wave quarks and gluons
are not a useful starting point for low-energy, confinement-dominated
physics}.  

    Of course the {\it full} story is contained in the elementary QCD
Lagrangian, and even in this situation it can be solved numerically using lattice QCD.
However, 
as always in physics, we need to understand the relevant degrees of freedom and the
effective forces between them for the phenomena under study if we wish to {\it understand}
QCD and be able to {\it use} our understanding to envision new phenomena and to extend the
power of QCD beyond our very limited ability to perform exact calculations using
the lattice.

%%%%%%%%%%%%%%%%%%%%%%%%%%%%%%%%%%%%%%%%%%%%%%%%%%%%%%%%%%%%%%%%%%
\section {All Roads Lead to Valence Constituent Quarks and Flux Tubes as the Dominant
Low Energy Degrees of Freedom}
%%%%%%%%%%%%%%%%%%%%%%%%%%%%%%%%%%%%%%%%%%%%%%%%%%%%%%%%%%%%%%%%%%

    From the preceeding discussion on the inappropriateness of the
perturbative quark and gluon degrees of freedom for describing the
phenomena of strong QCD, it will come as no surprise that
foremost among the puzzles we face is 
in fact a glaring  ``degree of freedom''
problem: the established low energy spectrum of QCD behaves as though it
is built from the degrees of freedom of spin-$1 \over 2$ fermions 
confined to a $q \bar q$ or $qqq$ system.  
Thus, for mesons of various flavors
we seem to observe   ``quarkonia"  spectra, while for  
the baryons  we seem to observe the 
spectra of the two relative coordinates of three spin-$1 \over 2$
 degrees of 
freedom.

    In particular, the empirical $N^*$ spectrum seems to demand that we use
these degrees of freedom. From the perspective of the QCD Lagrangian, this is a highly
nontrivial observation. Where are the expected gluonic states? Where are the extra
$q \bar q$ (meson cloud) degrees of freedom that would naively be expected to be as
important as the valence degrees of freedom?

    Compelling insights into these questions have
come from four different directions and converged
on a simple picture in which the appropriate degrees of freedom for strong QCD are 
valence {\it constituent} quarks and flux tubes. This picture is commonly referred to as
the flux tube model.

%%%%%%%%%%%%%%%%%%%%%%%%%%%%%%%%%%%%%%%%%%%%%%%%%%%%%%%%%%%%%%%%%%
\subsection{Spontaneous Chiral Symmetry Breaking in QCD}
%%%%%%%%%%%%%%%%%%%%%%%%%%%%%%%%%%%%%%%%%%%%%%%%%%%%%%%%%%%%%%%%%%

    Long before QCD was discovered, it was appreciated that the
underlying theory of the strong interactions should have a chiral
symmetry which is spontaneously broken \cite{NambuGoldstone}. When 
chiral symmetry is 
spontaneously broken, QCD must, by the Goldstone mechanism, 
generate the octet of light
Goldstone mesons like the pion. 
{\it Note that the forces which generate spontaneous chiral 
symmetry breaking 
and these mesons are not mysterious new forces: they are just 
the full $q
\bar q$ gluonic forces of QCD.} These forces give the vacuum nonperturbative
structure, in particular a nonzero value for the vacuum expectation 
value $\langle 0 \vert \bar q(x)  q(x) \vert 0 \rangle$ which leads directly
to a shift of the low energy effective quark mass from near zero to around
the standard constituent quark mass of 330 MeV. 
Given this,
spontaneous symmetry breaking  must also be associated with 
some effective spin-dependent force which splits the
$\pi$ from the $\rho$, the $K$ from the $K^*$, and the $\eta_{s \bar s}$ 
from the $\phi$. The venerable NJL model \cite{NJL} provides 
a prototypical example of how all of this happens.
Very roughly
speaking, the dynamics (which spontaneously breaks chiral symmetry
by creating  the $q \bar q$ condensate and thereby the constituent 
quark mass)
``conspire" in the pseudoscalar channel, with 
attractive forces compensating exactly (in the limit that the light 
quark Lagrangian masses are
zero) for the two constituent quark masses. 
In QCD, the role of the unspecified NJL dynamics must of course be 
played by gluonic forces,
though one can entertain various options for the nature of
these interactions. I shall return to a much more extensive discussion of these points and
the related problem of the mass of the $\eta '$ (the ``$U_A(1)$ problem"
\cite{UA(1)Problem}) below.

%%%%%%%%%%%%%%%%%%%%%%%%%%%%%%%%%%%%%%%%%%%%%%%%%%%%%%%%%%%%%%%%%%
\subsection{The Large $N_c$ Limit of QCD}
%%%%%%%%%%%%%%%%%%%%%%%%%%%%%%%%%%%%%%%%%%%%%%%%%%%%%%%%%%%%%%%%%%

It is now widely
appreciated that many of the observed
features of the strong interactions can
be understood in QCD within the $1/N_c$ 
expansion~\cite{LargeNc}. Moreover, there is growing
evidence from lattice QCD that  the main qualitative features
of QCD  are independent of $N_c$.  
This limit also provides the only known field theoretic basis for
the success of not only the valence quark model, but also of Regge
phenomenology, the narrow
resonance approximation, and  many of the systematics
of hadronic spectra and matrix elements. It can be shown in the large $N_c$ limit  that 
hadron two-point functions are dominated by graphs 
in which the valence quark lines propagate from their 
point of creation to their point of annihilation without 
additional quark loops.   A form of the OZI rule
also emerges naturally. Large $N_c$ QCD thus presents a picture
of narrow resonances interacting weakly with  hadronic 
continua. In this picture each resonance is
made of the valence  quarks of the quark model and glue {\it summed to all orders in} $g$.

%%%%%%%%%%%%%%%%%%%%%%%%%%%%%%%%%%%%%%%%%%%%%%%%%%%%%%%%%%%%%%%%%%
\subsection{Quenched QCD}
%%%%%%%%%%%%%%%%%%%%%%%%%%%%%%%%%%%%%%%%%%%%%%%%%%%%%%%%%%%%%%%%%%

    Quenched lattice QCD provides other new insights into 
QCD.
In quenched QCD the lattice sums amplitudes over all time histories
in which no $q \bar q$ loops are present. It thus gives quantitative
results  from an approximation with many elements in common with the
large $N_c$ limit. One of the most remarkable features of these
calculations is that despite what would seem to be a drastic approximation,
they
provide a reasonably good  description of low energy phenomenology. Indeed, for various
intermediate quantities like the QCD string 
tension they provide very good
approximations to full QCD results with the true lattice 
coupling constant replaced by an effective one. In 
quenched QCD, as in the large $N_c$ limit, two point functions
thus  seem to be well-approximated  by their valence quark content.

%%%%%%%%%%%%%%%%%%%%%%%%%%%%%%%%%%%%%%%%%%%%%%%%%%%%%%%%%%%%%%%%%%
\subsection{Some Comments on the Relation between the Lessons of the Large $N_c$ Limit
and of Quenched QCD}
%%%%%%%%%%%%%%%%%%%%%%%%%%%%%%%%%%%%%%%%%%%%%%%%%%%%%%%%%%%%%%%%%%

  In comparing the
large $N_c$ limit and quenched lattice QCD we note that:

\medskip

\noindent{$\bullet$}    In both pictures all resonances have only valence quarks,
but they have an unlimited number of gluons. Thus 
they support valence  models   for mesons  and 
baryons, but not for glueballs or for the gluonic content
of mesons  and 
baryons. 

\noindent{$\bullet$}  While both pictures tell us that hadrons are dominated by their
valence quark structure, the valence quark propagators are {\it not} nonrelativistic
propagators. I will elaborate on this point below.

\noindent{$\bullet$}   The large $N_c$ and quenched approximations are {\it not}
identical. For example, the $NN$ interaction is a $1/N_c$
effect, but it is not apparently suppressed in the quenched
approximation.

%%%%%%%%%%%%%%%%%%%%%%%%%%%%%%%%%%%%%%%%%%%%%%%%%%%%%%%%%%%%%%%%%%
\subsection{The Heavy Quark Limit}
%%%%%%%%%%%%%%%%%%%%%%%%%%%%%%%%%%%%%%%%%%%%%%%%%%%%%%%%%%%%%%%%%%

\begin{figure}
\begin{center}
\hskip -0.5cm
\epsfxsize=1.1in 
\epsfbox{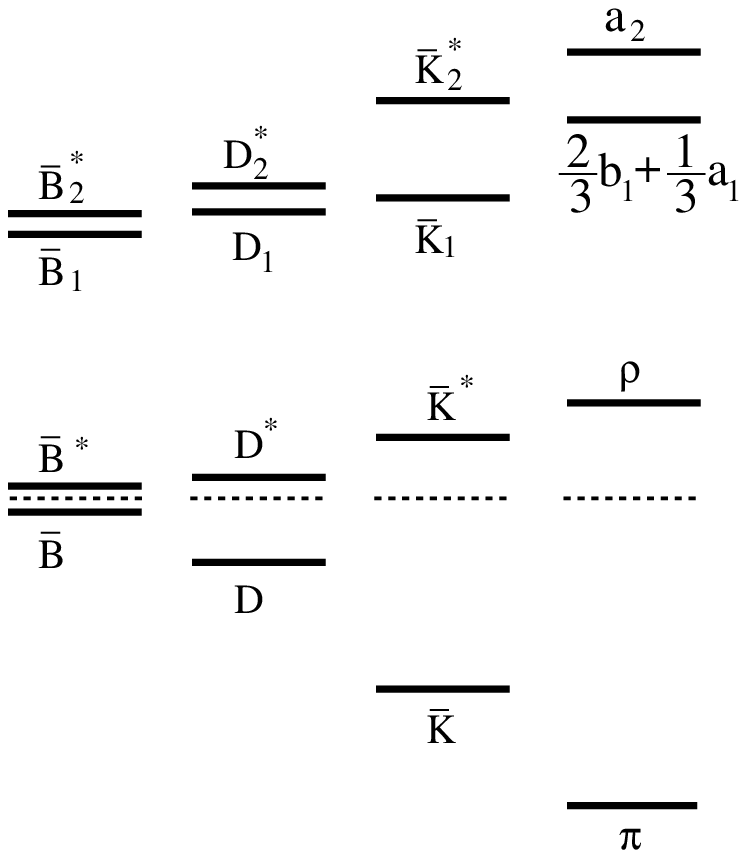}
\end{center}
\caption{The $Q \bar q$ 
meson spectra as a function of the ``heavy" quark mass.}
\label{fig:Qbarq}
\end{figure}

    The fourth perspective from which there is support for the same
picture is the heavy quark 
limit~\cite{IsgurWise}. While this limit has the weakest
theoretical connections to the light quark world, it has powerful 
phenomenological connections: see Fig. \ref{fig:Qbarq}. We see from this 
picture that in mesons containing a single heavy quark, 
$\Delta E_{orbital}$  (the gap between, for example, 
the $J^{PC}=1^{--}$
and $2^{++}$ states), is approximately independent of $m_Q$ while $\Delta E_{hyperfine}$
varies like $m_Q^{-1}$, both as expected in the heavy quark limit.

%%%%%%%%%%%%%%%%%%%%%%%

%%%%%%%%%%%%%%%%%%%%%%%

   Recall that in the heavy quark 
limit a hadronic two-point
function is dominated by a single valence $Q$ plus its associated
``brown muck", with neither $Q \bar Q$ loops nor $Q$ Z-graphs. The
fact that heavy-quark-like behaviour persists all the way down 
to light quark masses suggests that light quarks, like heavy quarks, 
behave like single valence quarks and thus by
extension that the ``brown muck" behaves like a single valence
antiquark.

%%%%%%%%%%%%%%%%%%%%%%%%%%%%%%%%%%%%%%%%%%%%%%%%%%%%%%%%%%%%%%%%%%
\section {Comments on the Degree of Freedom Problem}
%%%%%%%%%%%%%%%%%%%%%%%%%%%%%%%%%%%%%%%%%%%%%%%%%%%%%%%%%%%%%%%%%%

%%%%%%%%%%%%%%%%%%%%%%%%%%%%%%%%%%%%%%%%%%%%%%%%%%%%%%%%%%%%%%%%%%
\subsection {Where is the Glue and Why Represent it by a Flux Tube?}
%%%%%%%%%%%%%%%%%%%%%%%%%%%%%%%%%%%%%%%%%%%%%%%%%%%%%%%%%%%%%%%%%%

%%%%%%%%%%%%%%%%%%%%%%%
\begin{figure}
  \begin{center}
~
\epsfxsize=3.15cm \epsfbox{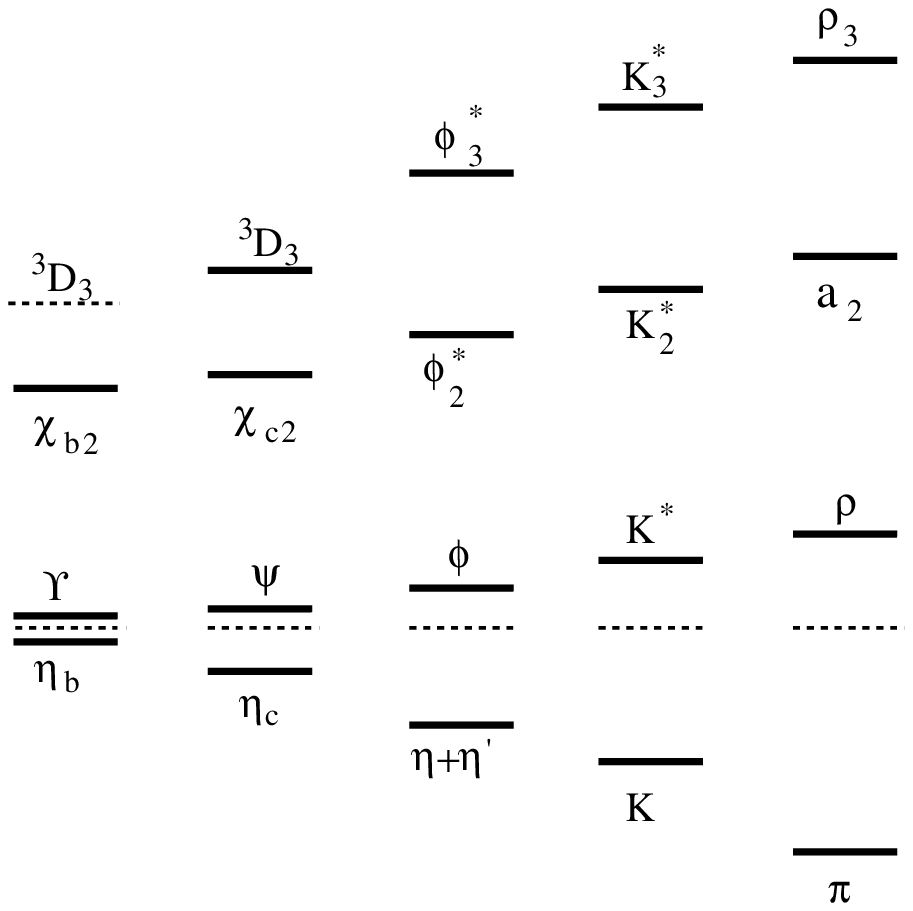}
\caption[x]{The $Q \bar Q$ 
meson spectra as a function of the ``heavy" quark mass.}
~
   \end{center}
\label{fig:QbarQ}
\end{figure}
%%%%%%%%%%%%%%%%%%%%%%%

   Fig. 2 shows that heavy quarkonium ($Q \bar Q$) behaviour apparently 
persists into the light meson spectrum. This is
surprising since the former spectrum depends on the decoupling of
gluonic excitations (as opposed to glue) from the spectrum. The flux
tube model \cite{IsgPat} provides an attractive explanation of the
adiabatic approximation which leads to this decoupling and which is now receiving
strong support from lattice QCD studies.  
Most importantly, the  
strongly collimated chromoelectric flux lines of the flux tube model have  been
seen on the lattice, as have the predicted first excited adiabatic surfaces of the flux
tube with an energy gap $\delta V(r)=\pi / r$ above
the quarkonium potential and
doubly-degenerate phonon quantum numbers.  This
basically requires  that the $J^{PC}$ exotic hybrid mesons predicted fifteen years
ago~\cite{IsgPat} exist.

   The flux tube model thus offers a possible explanation for one of the most puzzling
apparent inconsistencies between the valence quark model and QCD. Moreover,
as  discussed above, in the large $N_c$ limit of QCD hadrons do indeed
consist of just their valence quarks and the glue between them.  Thus the flux
tube model may  be viewed as a 
realization of QCD in the large  $N_c$
limit which is in addition consistent with
insights into strong QCD which have emerged from spontaneous chiral symmetry breaking, the
large
$N_c$ limit, from quenched lattice QCD, and from heavy quark theory. It corresponds to a
rigorous and in principle systematic expansion of QCD around the strong coupling limit
rather than the weak coupling limit \cite{IsgPat}. 

    We should  remember that many of these theoretical developments
have their historical roots in data. In particular, the leading Regge trajectories
($1^-$, $2^+$, $3^-$, $4^+$, $5^-$, $6^+$,  etc. in mesons,
$1/2^+$, $3/2^-$, $5/2^+$, $7/2^-$, $9/2^+$, $11/2^-$ etc. in the
$N^*$'s, and $3/2^+$, $5/2^-$, $7/2^+$, $9/2^-$, $11/2^+$, $13/2^-$ etc. in the
$\Delta^*$'s) led Nambu \cite{NambuString} to postulate in the early 1970's that QCD was a
relativistic string theory with the mass of states at high excitation dominated by the 
mass of the string. In QCD, Nambu's prophetic picture is realized via quarks confined
to infinite towers of (narrow) bound states by a string-like flux tube.

    Of course, more data is
required to confirm that the flux tube model is  indeed on the right path and to
define the next steps down this path. This is the goal of  Hall D of the 12 GeV Upgrade
project, but there are very important implications as well for the current $N^*$
program: it has recently become clear~\cite{NIonHybridCrossSections} that this picture
also requires that the cross sections for hybrid $N^*$ electroproduction be
large.

%%%%%%%%%%%%%%%%%%%%%%%%%%%%%%%%%%%%%%%%%%%%%%%%%%%%%%%%%%%%%%%%%%
\subsection {Constituent Quarks are Not Nonrelativistic Quarks}
%%%%%%%%%%%%%%%%%%%%%%%%%%%%%%%%%%%%%%%%%%%%%%%%%%%%%%%%%%%%%%%%%%

   Discussions of the constituent quark model are sometimes confused by a  failure to
recognize that the  nonrelativistic quark model provides
a very convenient qualitative tool for understanding many of the features of strong QCD,
but should not be interpreted literally. In particular,
all of the QCD-based pictures described above have  propagating valence quarks with
contributions from not only a positive energy quark propagator, but also from 
 ``Z-graphs".  (A ``Z-graph" is a time-ordered graph in which 
the interactions first produce a pair and then 
annihilate the antiparticle of the produced pair 
against the original propagating particle).  Cutting 
through a  two-point function at a fixed time 
therefore would in general reveal not only the valence 
quarks but also a large $q \bar q$ sea.
This  does not seem to correspond to the usual 
valence approximation.
Consider, however, the Dirac equation for a single light 
quark interacting with a static color source.  This equation represents 
the sum of a set of Feynman graphs which also include 
Z-graphs, but the effects of those graphs is captured 
in the lower components of the single particle Dirac 
spinor. {\it I.e.}, such Z-graphs correspond to relativistic 
corrections to the quark model.  That such corrections 
are important has been known for a 
long time.  For us the relevant point is that while 
such effects have important quantitative effects on quark model 
predictions ({\it e.g.}, they are commonly held to be 
responsible for much of the required reduction of 
the nonrelativistic quark model prediction  
that $g_A=5/3$ in neutron beta decay), they do not qualitatively 
change the single-particle nature of the spectrum of 
the quark of our example, nor would they necessarily
qualitatively change the spectrum of $q \bar q$ or $qqq$ systems.

%%%%%%%%%%%%%%%%%%%%%%%%%%%%%%%%%%%%%%%%%%%%%%%%%%%%%%%%%%%%%%%%%%
\subsection {The Valence Approximation is an Approximation}
%%%%%%%%%%%%%%%%%%%%%%%%%%%%%%%%%%%%%%%%%%%%%%%%%%%%%%%%%%%%%%%%%%

    The valence approximation can only be taken as a starting point for a systematic
treatment of
strong QCD, since we know that $q \bar q$ pair creation plays an
important role in many phenomena. In particular, 

\medskip

\noindent $\bullet$  While particle systematics, including the linearly rising Regge
trajectories, require a narrow resonance approximation, we know that the resonances have
finite but nonzero widths
$\Gamma << M$.

\noindent $\bullet$ In analyzing data, we are clearly faced with the fact that the $N^*$'s
are immersed in a  baryon-meson continuum, and that the interplay between the resonance
spectra and these continua are important.

\noindent $\bullet$ Even amongst the more sophisticated analyses of the $N^*$ spectra,
there has historically been a great deal of confusion about the relationship between the
predictions of the valence quark model for spectra and the narrow resonance approximation.
The large $N_c$ limit has begun to clarify matters: while the quark model spectrum {\it
is} a narrow resonance approximation, it is a {\it mixture} of leading effects in the
$N_c$ expansion and $1/N_c$-suppressed effects. Thus the connection of quark model
predictions to the data is quite subtle, and cannot be treated by simply considering the
interaction of narrow resonances with the continua {\it via} meson loop graphs. For
example, many of the effects of meson loop graphs are already subsumed into the quark
model's string tension \cite{NIonThresholds}.

\noindent $\bullet$ The large $N_c$ limit tells us that the valence approximation will be
good to an accuracy of order $1/N_c$ {\it except for SU(3) singlet channels}
where the valence predictions will be multiplied by a correction factor of not
$(1+constant~N^{-1}_c)$, but rather $(1+constant~N_f N^{-1}_c)$, where $N_f$ is the number
of light flavors. Thus, for example, the proton spin crisis may be attributed to the
fact that the  valence spin contribution  is
reduced coherently by relatively small  effects from {\it each of 
$u \bar u$, $d \bar d$, and $s \bar s$ pairs} from the quark-antiquark sea. 

\medskip

\noindent While the above points make it clear that the quark-antiquark sea must be
added to the valence quarks to get an accurate picture of hadrons and their
interactions, naively attempting to add
$q \bar q$ pair creation to the valence quark model leads to a number of very serious
problems.  These problems and potential solutions 
to them have been extensively discussed in 
a series of papers on ``unquenching" the quark model
\cite{NIonThresholds}.  The most important lesson
of these studies is that low energy truncations of the tower of meson loops corrections to
the valence quark states (which generate the
$q \bar q$ sea; see Fig. 3) are usually misleading.

\begin{figure}
  \begin{centering}
  \setlength{\unitlength}{0.022cm}%
  \begin{picture}(404,220)(117,610)
  \thicklines
  \put(480,719){\oval( 80, 80)[br]}
  \put(480,719){\oval( 80, 80)[tr]}
  \put(480,719){\oval( 40, 40)[br]}
  \put(480,719){\oval( 40, 40)[tr]}
  \put(480,799){\line( 0,-1){ 40}}
  \put(480,739){\line( 0,-1){ 40}}
  \put(480,679){\line( 0,-1){ 40}}
  \put(460,799){\line( 0,-1){160}}
  \put(440,799){\line( 0,-1){160}}
  \put(208,754){\line( 2,-3){ 11.539}}
  \put(208,686){\line( 2, 3){ 11.539}}
  \put(181,761){\line( 6,-1){ 20.919}}
  \put(181,678){\line( 6, 1){ 20.919}}
  \put(222,730){\line( 0,-1){ 21}}
  \put(180,800){\line( 0,-1){160}}
  \put(117,762){$$ \large (a) $$}
  \put(375,760){$$ \large (b) $$}  
  \end{picture}
\label{fig:qbarqloop}
  \caption[x]{
  A meson loop correction to an $N^*$ propagator,
   at (a) the hadronic level, and (b) showing one graph it generates at the quark level.}
  \label{fig:loops}
  \end{centering}
\end{figure}
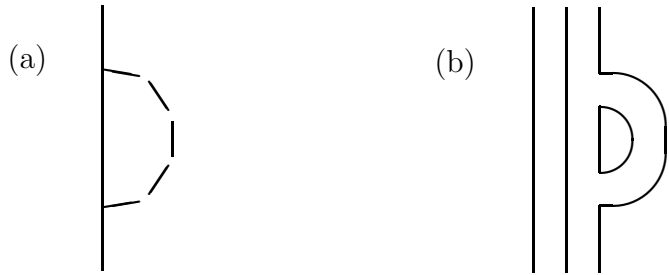 
\vspace{0.3cm}
%

%%%%%%%%%%%%%%%%%%%%%%%%%%%%%%%%%%%%%%%%%%%%%%%%%%%%%%%%%%%%%%%%%%
\section {What Are the Effective Forces between the
Low Energy Degrees of Freedom?}
%%%%%%%%%%%%%%%%%%%%%%%%%%%%%%%%%%%%%%%%%%%%%%%%%%%%%%%%%%%%%%%%%%

    There are  many contending pictures for the origin of
the effective forces between the low energy degrees of freedom of strong QCD.
I have observed, perhaps as a result of the variety of proposals, some confusion and 
unnecessary contentiousness over this key issue. In some cases there {\it are}
real issues to be resolved, but not always.

%%%%%%%%%%%%%%%%%%%%%%%%%%%%%%%%%%%%%%%%%%%%%%%%%%%%%%%%%%%%%%%%%%
\subsection {The Confining Interaction}
%%%%%%%%%%%%%%%%%%%%%%%%%%%%%%%%%%%%%%%%%%%%%%%%%%%%%%%%%%%%%%%%%%

    The confining interaction is automatically provided by the flux tube, and in
the very low energy regime where the adiabatic approximation is valid, the flux tube
degrees of freedom can be replaced by an adiabatic effective potential, at least for heavy
enough quarks.

    The central part of this potential is linear at large distances. This fact should not
be confused with discussions of the basis space for solutions to spectroscopic
problems, which is an issue of accuracy and not principle. For example, in the
Isgur-Karl model
\cite{IsgurKarl}, a harmonic basis is used to solve the linear potential problem. A
harmonic potential is {\it not} assumed.

     There have historically been many many proposals for the spin dependence of the
confining potential; very recently, there has been much discussion of its spin-orbit
components. 
The basic issues are  discussed in the original Isgur-Karl
papers,  since in that model all spin dependence is assumed to arise from one gluon
exchange (OGE) except for the unavoidable purely kinematic  spin-orbit force from
Thomas precession in the confining potential. The key point is that the central
confining potential, whose eigenstates almost every modeller takes as the basis for
spin-dependent perturbations of resonance spectra, {\it must}  produce very strong
spin-orbit forces through Thomas precession. To analyze the implications of this fact  it
is best to examine the meson sector.  The mesons appear to have a
``spin-orbit  problem" as can be seen by examining the first band of positive parity
excited mesons:  the four P-wave mesons of every flavor are nearly degenerate.  Most of
the observed small non-degeneracies are due to hyperfine interactions, but the spin-orbit
matrix elements can be extracted.  For example, by taking the isovector meson combination 
${5 \over 12}m_{a_2}-{1 \over 4}m_{a_1}-{1 \over 6}m_{a_0}$ 
one can isolate their spin-orbit matrix
element of $-3 \pm 20$ MeV.  This matrix element
is much smaller than would be obtained from either the standard short-distance OGE
interaction or from the  Thomas precession term alone. However,  the large ``normal"
OGE spin-orbit matrix element  tends  to cancel the strong
``inverted" spin-orbit matrix element from Thomas precession. Typical fits to the data,
including the heavy quarkonia, give light quarkonium OGE and Thomas precession spin-orbit
matrix elements of about $\pm 200$ MeV, respectively.

	The physics behind this cancellation has received support recently
from the theory of $Q \bar Q$ systems, where both analytic techniques
and numerical studies using lattice QCD 
have shown that the confining forces {\it are} spin-independent  apart from
the inevitable spin-orbit pseudoforce due to Thomas precession.  Moreover,
the data on charmonia  require the
inverted spin-orbit matrix element from Thomas precession in the confining potential
to cancel part of
the strength of their rigorously required positive OGE matrix element.  If the charm quark
were more massive, its low-lying spectrum would be totally dominated
by one gluon exchange.
Indeed, one observes that the $\Upsilon$ system is closer to this ideal, as
expected.  Conversely, as one moves from $c \bar c$ to lighter quarks, the $\ell=1$
wave functions move farther out into the confining potential and the
relative strength of the Thomas precession term grows.  It is thus very
natural to expect a strong cancellation in light quark systems, though from this
perspective the observed nearly perfect cancellation must be viewed as  accidental.

     Given this rather satisfactory solution of the ``meson spin-orbit problem", it would
be very surprising if the analogous ``baryon spin-orbit problem" didn't have an analogous
solution, so this assumption was built into the Isgur-Karl model. I will return to a
discussion of these matters below.

    To summarize: there is very strong evidence that confinement may be approximated at
low energy by replacing the flux tube with an effective potential with a linear central
component. In the leading correction to this nonrelativistic approximation, the potential
should be supplemented with {\it only} the minimally required Thomas precession term
(which acting alone would produce strongly inverted spin-orbit multiplets) and with
spin-independent but momentum-dependent terms which take into account the impact of the
orbital angular momentum of the massive rotating flux tube on the spectrum of states.

%%%%%%%%%%%%%%%%%%%%%%%%%%%%%%%%%%%%%%%%%%%%%%%%%%%%%%%%%%%%%%%%%%
\subsection {The Short-Distance Interactions}
%%%%%%%%%%%%%%%%%%%%%%%%%%%%%%%%%%%%%%%%%%%%%%%%%%%%%%%%%%%%%%%%%%

%%%%%%%%%%%%%%%%%%%%%%%%%%%%%%%%%%%%%%%%%%%%%%%%%%%%%%%%%%%%%%%%%%
\subsubsection {One Gluon Exchange}
%%%%%%%%%%%%%%%%%%%%%%%%%%%%%%%%%%%%%%%%%%%%%%%%%%%%%%%%%%%%%%%%%%

    The preceeding discussion leaves  little room for doubt about the confining
interaction and its structure, at least to leading order. There is, however, room
for doubt about the nature of the short-distance interactions, and in particular whether
the standard OGE model for these interactions is correct. While  healthy, this
discussion has also sometimes been quite confused, and I want to take this opportunity to
try to sort out the issues. Let me acknowledge from the beginning that I am a partisan of
the OGE school, and so warn you that my organization of the issues (but not my facts!)
may be prejudicial.

    The discussion is once again best started in the simpler meson sector.
Figure 2 shows what we
know about the evolution of quarkonium spectroscopy as a function of the quark masses.  In
heavy quarkonia ($b \bar b$ and
$c \bar c$) we {\it know} that hyperfine interactions are generated by OGE
perturbations of wave functions which are solutions of the Coulomb plus linear
potential problem.  I find it difficult to look at this diagram and not see
a smooth evolution of the wavefunction (characterized by the smooth evolution
of the orbital excitation energy) convoluted with the predicted $1/m_Q^2$
strength of the OGE hyperfine interaction. OGE-based quark models turn this qualitative
impression into quite good quantitative predictions.

	This same conclusion can be reached by approaching the light
quarkonia from another angle.  Figure \ref{fig:Qbarq}                                
shows the
evolution of heavy-light meson hyperfine interactions from the heavy quark limit to the
same isovector quarkonia.  In this case we {\it know} that in the heavy quark limit
\cite{IsgurWise} the hyperfine interaction is given by the matrix element of the operator
$\vec \sigma_Q \cdot \vec B/2m_Q$.
The conclusion that heavy-light meson hyperfine interactions are
controlled by OGE is thus also supported by the striking $1/m_Q$ behaviour of the
ground state splittings in Fig. \ref{fig:Qbarq}
as $m_Q$ is decreased from $m_b$ to $m_c$ (in which systems it may be rigorously applied)
on down to $m_s$ and then to
$m_d$: it certainly appears that for all quark masses the quark Q interacts
with $\vec B$ through its chromomagnetic moment $\vec \sigma_Q/2m_Q$, as would be
characteristic of the OGE mechanism.

%%%%%%%%%%%%%%%%%%%%%%%
\begin{figure}
  \begin{center}
~
\epsfxsize=7cm \epsfbox{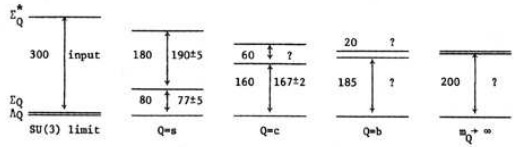}
\caption[x]{Ground state  baryon  hyperfine splittings in heavy-light
systems as a function of the mass $m_Q$ of the heavy quark. The spectra on the far right
are the $m_Q \rightarrow \infty$ limits of heavy quark symmetry. The  
$\Sigma^*_Q-\Lambda_Q$ splitting 
and the 
positions of 
$\Sigma^*_b$ and $\Sigma_b$ 
are estimates from the quark model (for which there is now experimental support); all other
masses are from experiment.  The spectra are shown to scale and
may conveniently be calibrated with the  $\Sigma_c-\Lambda_c$ splitting of 169 MeV.}
~
   \end{center}
\label{fig:udQ}
\end{figure}
%%%%%%%%%%%%%%%%%%%%%%%

     Let us now turn to baryons. As shown in Fig. \ref{fig:udQ}, the baryon analog of
Fig. \ref{fig:Qbarq}, experiment provides further strong 
evidence in support of the dominance of OGE in the baryons. 
It is clear from this Figure that in the heavy quark
limit the 
splittings are once again behaving like $1/m_Q$ as demanded by heavy quark
theory, and  once again it is
difficult to look at this diagram and not see a smooth evolution of this $1/m_Q$
behaviour from $m_b$ to $m_c$ to $m_s$ to $m_d$, where by SU(3) symmetry
$\Sigma^*_{SU(3)} - \Lambda_{SU(3)}=\Delta - N$, the splitting relevant to $N^*$'s. 
Indeed, using standard constituent quark masses  ($m_d=m_u \equiv m =0.33$ GeV, $m_s=0.55$
GeV,
$m_c=1.82$ GeV, $m_b=5.20$ GeV), the OGE mechanism with its natural $1/m_Q$ behaviour
{\it quantitatively} describes these spectra. Thus one  observes in both mesons
and baryons the remnants  of heavy quark spectroscopy in light
quark systems.  For example, in Fig. \ref{fig:Qbarq} the $D^*-D$ heavy quark spin
multiplet is naturally identified with the $K^*-K$ multiplet, {\it i.e.,} the basic
degrees of freedom seen in the spectrum are the same, and from the observed $D^*-D$
splitting and the $1/m_Q$ heavy quark scaling law alone one expects  a $K^*-K$ splitting of 460 MeV,
quite close to the actual splitting of 400 MeV  (including wave function effects
improves the prediction).
One might try to escape this conclusion by arguing that
between $m_c$ and $m_s$ the OGE-driven $1/m_Q$ mechanism turns off 
and some other
mechanism which imitates it turns on. I do not know how to rule out this
baroque possibility except on a case-by-case basis.

    The excited
charmed baryon sector has recently provided further strong evidence for
the dominance of the OGE mechanism in baryons. From an analysis of their decay
patterns, it seems very likely that {\it if}
the $\Lambda (1405){1 \over 2}^-$ and $\Lambda (1520){3 \over 2}^-$ are
valence-quark-model-type $uds$ states, then they are the expected spin-orbit partners of
the quark model. Heavy quark symmetry \cite{IsgurWise} demands that in the heavy-light
isospin zero $\Lambda_Q$ sector, the $\Lambda_Q{1 \over 2}^-$ and $\Lambda_Q {3 \over
2}^-$ be degenerate as $m_Q \rightarrow \infty$ and that their splitting open up like
$1/m_Q$ as $m_Q$ decreases.  The $\Lambda_c(2594){1 \over 2}^-$ and 
$\Lambda_c(2627){3 \over 2}^-$ 
appear to be such a nearly degenerate pair of states in the charmed baryon
sector.  Their center-of-gravity 
${2 \over 3}m_{\Lambda_Q {3 \over 2}^-}  + {1 \over 3}m_{\Lambda_Q{1 \over 2}^-}$
is 330 MeV above the
$\Lambda_c(2285)$.  This is to be compared with the center-of-gravity of the
$\Lambda (1520)$ and $\Lambda (1405)$ which lies 365 MeV above the $\Lambda(1115)$, 
in accord with the expectation  that the
orbital excitation energy of the negative parity excitations of $\Lambda_Q$
will be a slowly increasing function of $1/m_Q$.  (Recall Fig. \ref{fig:Qbarq}: the same
is true in mesons.)   This alone suggests that the strange quark analogs of the heavy
quark spin multiplet 
$(\Lambda_c(2627){3 \over 2}^-, \Lambda_c (2594){1 \over 2}^-)$  should exist just around
the mass of the $\Lambda (1520){3 \over 2}^-$ and $\Lambda (1405){1 \over 2}^-$.    
Moreover, using the
predicted $1/m_Q$ behaviour of the  multiplet
splitting, one would expect a splitting of  $\simeq 110$ MeV in
the
$\Lambda$ sector compared to the observed splitting of 115 MeV. It is thus 
difficult to avoid identifying 
$(\Lambda (1520){3 \over 2}^-, \Lambda (1405){1 \over 2}^-)$ as
the strange quark analogs of a heavy quark spin multiplet, and
to avoid concluding that the $1/m_Q$ evolution of the OGE mechanism is responsible
for its splitting.

   Of course in both Figs. \ref{fig:Qbarq} and \ref{fig:udQ}, the extrapolation of the
nonrelativistic OGE mechanism to light constituent quark masses cannot be taken
literally. In $Q \bar Q$ systems, OGE is treated by using a nonrelativistic
reduction to define a low energy effective potential; such a treatment
would not be accurate for light quarks whose masses are of the order of the QCD scale.
What quenched lattice QCD and the large $N_c$ limit
tell us is that mesons and baryons are dominated by their valence quark structure, 
and what Figs. \ref{fig:Qbarq} and \ref{fig:udQ} strongly suggest is that a {\it smooth}
extension of the $Q \bar Q$  valence quark interaction is in operation  for all quark
masses. For light quark masses this extended OGE interaction will include many new
effects, including not only straightforward relativistic corrections, but also those like
Z-graphs which arise because  the light valence quarks are embedded in a relativistic
field theory and so have instantaneous projections into subspaces with additional $q \bar
q$ pairs, and effects of vacuum  structure that are not important at short distances.

    This brings us to one of the most confused elements of the discussions of the quark
model, that it is somehow inconsistent with chiral dynamics. This is certainly not true
in principle. As discussed in Section 3.1, spontaneous chiral symmetry breaking generates
the constituent quark mass, so this element of quark models is {\it required} by chiral
dynamics. Moreover, the pion is not a magical massless Goldstone boson which appears out
of nowhere and is thus outside of the domain of quark models, in stark contrast to the
discussion (and the experimental data!) above. A massless
pion automatically appears in both  quenched lattice QCD and the large $N_c$ limit as the
mass of the lightest state of a propagating quark-antiquark pair. I.e., even in the 
valence quark approximation, a massless pion is automatic!

%%%%%%%%%%%%%%%%%%%%%%%%%%%%%%%%%%%%%%%%%%%%%%%%%%%%%%%%%%%%%%%%%%
\subsubsection {Instantons}
%%%%%%%%%%%%%%%%%%%%%%%%%%%%%%%%%%%%%%%%%%%%%%%%%%%%%%%%%%%%%%%%%%

    There is another possibility for the short-distance interactions of light quarks which 
has deservedly been receiving a great
deal of attention recently: the
instanton liquid model 
\cite{Shuryak}. This picture is inconsistent with {\it any} model based on a
Feynman-diagrammatic treatment of QCD (even  all-orders ones like the large $N_c$ limit).

    Let me begin with a little history. For many years, QCD was plagued by
the ``$U_A(1)$
problem" \cite{UA(1)Problem}:  the equations of motion of QCD imply that spontaneous
chiral symmetry breaking leads
to {\it nine} and not just eight Goldstone bosons, a conclusion
that is apparently inconsistent with the large mass of the $\eta '$ meson at about 1 GeV.
However, the
$U_A (1)$ current is anomalous, and by the late 1970's it was understood through the
study of instantons
\cite{instantons,tHooftInstantons} that the anomaly leads to a
nonconservation of the $U_A (1)$
charge and thereby to the evasion of Goldstone's theorem when chiral
symmetry is spontaneously
broken.  The  connection between the quark model  and
instantons was discussed
by Witten \cite{WittenUA(1)}, Veneziano \cite{Veneziano}, and others, who explored
more generally the conflict between instantons and
the large $N_c$ expansion.
It is certainly a mainstream, viable,  and widely  held
belief that the $U_A(1)$ problem is solved through instanton
contributions to the axial anomaly.  However, as emphasized by Witten,
instantons vanish like $e^{-N_c}$ and so do not appear in the large $N_c$
expansion.  ``Insofar as [instantons play] a significant role in the
strong interactions, the large $N_c$ expansion must be bad.  It is
necessary to choose between the two."

      What is the essential origin of this conflict? Unlike QED, the nonabelian structure
of QCD admits many possible classical vacua characterized by a ``winding number" which
classifies the topology of the nonabelian vector potential, but all of which have the
zero color electric and magnetic fields and thus zero energy. (In QED there appear to be
many choices for a vector potential which have zero  electric and magnetic fields
and zero energy, but they are all related by  gauge transformations and are thus
equivalent. A gauge transformation cannot change winding number (it is a
topological property), and thus the infinite number of classical QCD vacua are
distinct.) The instanton picture begins from the premise
that the correct  vacuum around which one should compute quantum corrections
(i.e., on which a normal Feynman diagram expansion should be made) is the one resulting
from  tunneling between these energy-equivalent classical vacua (much like the
tunneling between two potential wells separated by a classically forbidden region). When
the vacuum tunnels between two of its possible states (characterized by different winding
numbers), it does so quickly (by the uncertainty principle); such an event is called an
instanton. The tunneling process thus dynamically generates the true vacuum of QCD as a
superposition of classical vacua of various winding numbers. It turns out that this
superposition can be characterized by  single parameter called $\theta$.

    The physics of $\theta$ is a very important story all unto itself. If $\theta$ is
nonzero as naively expected from the previous discussion (it is after all a strong
interaction parmeter!), then the strong interactions would violate $CP$ invariance.
Experimental constraints tell us that $\theta$ is roughly ten orders of magnitude
smaller than we might have expected! This is known as the ``strong CP problem". Axions
were invented to try to deal with this problem, but there is no established solution.  At
the moment, $\theta$ must be listed as one of the 17 basic parameters of the standard
model (though its value is measured to be negligibly small).

    Let us leave this interesting digression to return to our main story. Since the usual
Feynman diagrammatic expansion of QCD is built on the perturbative vacuum with zero
vector potential, it obviously has no knowledge of other vacua or of the instantons that
connect them, so the effects of such instanton events  have to be added ``by hand" as
a supplementary effective interaction. For us the
relevant point is that  this effective interaction includes an interaction between
the quarks known as the 't Hooft interaction \cite{tHooftInstantons}.  As noted by
Witten \cite{WittenUA(1)},  if instantons are important, then  the large
$N_c$ expansion would fail, since it assumes that all-orders properties of the QCD
Feynman diagrammatic expansion are properties of QCD.   

    What would the impact be on the effective interaction we are trying to determine? The
instantons (which are still a purely
gluonic effect!) would require the addition of the 't Hooft interaction to the
quark model. This interesting possibility  would indicate that the Coulomb plus linear
picture,  when extrapolated 
to light quarks, will fail {\it quantitatively} and that light quark 
spin-dependence, while still
of gluonic origin in both mesons and baryons, will in general
be a mixture in some proportions of OGE-type forces and the `t Hooft interaction.
However, as Witten argued
long ago \cite{WittenUA(1)}, given the $U_A(1)$ anomaly and confinement,
the large
$N_c$ limit is also able to explain the $\eta'$ mass
{\it without} instantons: large confinement-generated gluonic fluctuations in the vacuum
also couple to the $U_A(1)$ anomaly and change the winding number. No matter which solution
to the
$\eta'$ mass is correct, there is still $\theta$-dependence in QCD and thus a strong $CP$
problem. But for strong interaction physics, understanding whether instantons are or
are not important remains one of the most important outstanding issues \cite{IsgurThacker}.

%%%%%%%%%%%%%%%%%%%%%%%%%%%%%%%%%%%%%%%%%%%%%%%%%%%%%%%%%%%%%%%%%%
\subsubsection {The Pion Exchange Model for Interquark Forces}
%%%%%%%%%%%%%%%%%%%%%%%%%%%%%%%%%%%%%%%%%%%%%%%%%%%%%%%%%%%%%%%%%%

    An alternative to the OGE model for the short-distance forces between quarks is the
pion exchange model (OPE model), which posits that these forces are dominated by one
pion exchange between the constituent quarks \cite{GR}. More generally, the model assumes
that the short-distance forces between the quarks in a baryon is mediated by the exchange
of the octet of nearly massless Goldstone bosons generated by the spontaneous breaking of
chiral symmetry. It is argued that the quark model requires
such effects to be consistent the spontaneous breaking of  chiral symmetry, and
that the replacement of OGE by OPE solves many of the problems of the quark model for
baryons. I have explained above why there is no need to invoke a
special mechanism like OPE to make the OGE model  consistent with the spontaneous breaking
of with chiral symmetry. I have also presented my objections to the 
OPE model in  conference talks. More recently, I have
published these criticisms \cite{NIonOPE}, taking into account
recent attempts to overcome some of the problems of the original OPE model by extending
it to include heavier meson exchanges and other effects.  I  briefly summarize the
main points here:

\medskip

$\bullet$  One of the original motivations for the OPE model was that it could solve
the baryon spin-orbit problem since OPE, in contrast to OGE, produces no spin-orbit force.
From the discussion of Section 5.1 on the inverted spin-orbit potential generated by
confinement, one can see that this motivation was based on a misunderstanding of the
nature of the problem, which is to arrange a sufficiently precise cancellation between
the inevitable Thomas precession term generated by confinement and short-range
dynamical spin-orbit forces. Recent elaborations of the original OPE model to include
the exchange between quarks of higher mass mesons can produce such dynamical spin-orbit
forces, but in these circumstances this model clearly also has a ``spin-orbit problem",
i.e., no advantage has been achieved. On the contrary, since higher mass mesons will
have only short-range effects, it may be very difficult to  arrange  a
cancellation of spin-orbit forces in both the $L=1$ and $L=2$ baryons as
required by the data.

$\bullet$   Another of the original claims of the OPE model is that it produces
a superior description of baryon masses. I will comment briefly on this
somewhat complicated claim below, but in Ref. 19, I focus on a more straightforward
matter: in a complex system like the baryon resonances, predicting the spectrum of states
is not a very stringent test of a model.  The prototypical example (and the first case in
$N^*$ spectroscopy where this issue arises) is the two $N^*{1\over 2}^-$ states found in
the 1500-1700 MeV range. In any reasonable valence quark model, two $N^*{1\over 2}^-$
states will be predicted in this mass range since totally antisymmetric states
with overall angular momentum  ${1 \over 2}$ can be formed
by coupling  either quark spin ${3 \over 2}$ or quark spin ${1 \over 2}$ with $\ell=1$.  
In the
general case such a model will therefore give
\begin {eqnarray}
\vert N^*{1\over 2}^- (upper) \rangle &=& cos \theta_{{1\over 2}^-} \vert ^4P_N \rangle
+sin \theta_{{1\over 2}^-} \vert ^2P_N \rangle \\
\vert N^*{1\over 2}^- (lower) \rangle &=& cos \theta_{{1\over 2}^-} \vert ^2P_N \rangle
-sin \theta_{{1\over 2}^-} \vert ^4P_N \rangle
\label{eq:nhalfmixing}
\end {eqnarray}
in an obvious notation.  Since the masses of these resonances are only
known (and currently interpretable) to roughly 50 MeV, it is not very difficult
to arrange for a model to give a satisfactory description of the $N^*{1\over 2}^-$
spectrum.  However, among models which ``perfectly" describe the spectrum
there is still a continuous infinity of predictions for the internal
composition of these two states since all values of $\theta_{{1\over 2}^-}$ from 0 to $\pi$
correspond to distinct states. The correct mixing angle has been known experimentally for
25 years. It is correctly predicted by the OGE model, but not by the OPE model.

	        This is the simplest example of examining the {\it complete} spectroscopy (and
not just the masses) of baryons. In the positive parity states, an analysis which fails to
examine the internal composition of states is nearly useless. For example, the valence
quark model  predicts {\it five}
$N^*{3\over 2}^+$ states, but only one is known.  It would be an unlucky modeller who
couldn't identify one of their five predicted states with the observed state and
claim success!

$\bullet$        There are two ways in which the OPE model is a disaster for mesons: since
it doesn't produce spin-dependent interactions in mesons (see below), it requires that
different mechanisms are dominant in these two systems despite the great similarity of
their spin-independent spectra (and the implied similarity of their internal structure),
and it predicts the existence of effects in mesons which are ruled out experimentally.

     We know from quenched lattice QCD that the bulk of both
meson and baryon hyperfine  interactions occur in the quenched
approximation  ({\it i.e.,} in the absence of closed $q \bar q$ loops).  
There are nevertheless Z-graph-induced meson exchanges between quarks that
arise in this approximation that could in principle be 
the origin of those posited in OPE-type models. However, such
meson exchange can only operate between two quarks and  {\it not}
between a quark and an antiquark, with the unsatisfactory consequence just mentioned.

    The second problem in mesons is that the OPE model would produce unacceptably
large violations of the OZI rule.  While meson exchange cannot produce interactions
between the quark and antiquark in a meson, it would lead to $q \bar q$ annihilation
which will badly violate the OZI rule.

$\bullet$  The OPE model is not consistent with the heavy quark symmetry of QCD
for $m_Q \rightarrow \infty$. Recall the discussion associated with Fig. 4.
It is clear from these spectra that in the heavy quark
limit the OPE mechanism is not dominant:  exchange of the heavy
pseudoscalar meson $P_Q$ would produce a hyperfine interaction that scales
with heavy quark mass like $1/m_Q^2$, while for heavy-light baryons the
splittings are behaving like $1/m_Q$ as in the heavy-light mesons.

\medskip

    Ref. 19 does not discuss the claim that the OPE models produce a better spectrum than
the Isgur-Karl model. I have explained that without using  mixing angle information,
spectroscopy is a rather blunt tool. It is for this reason that I did not engage this
argument in my paper, but I want to take this opportunity to clear up a 
misconception about the spectroscopy of the Isgur-Karl model that has
been labelled the ``level ordering problem". Begin in the
$S=0$ states where we have the $N$,
$N^*$(1535), and $N^*$(1440). In the harmonic limit, these states would be equally spaced
in the order they are listed instead of the order in which they are seen. Despite the
impression that the OPE papers might give, the  observed pattern {\it is
automatically} realized in the Isgur-Karl model.  The inversion is due to two effects.

The first is that the true confining potential will rigorously break a harmonic
oscillator spectrum (to leading order in the anharmonicity, which may be minimized by
choice of the unphysical oscillator constant of the basis space) into the following pattern
in terms of
$SU(6)$ multiplets (note that I have arranged these equations so that the lightest state
is at the bottom of the array):
\begin{eqnarray}
   E(20,1+)&=& E_0 + 2 \Omega    \\
   E(70,2+)&=& E_0 + 2 \Omega - 1/5 \Delta    \\
   E(56,2+)&=& E_0 + 2 \Omega - 2/5 \Delta    \\
   E(70,0+)&=& E_0 + 2 \Omega - 1/2 \Delta    \\ 
   E(56',0+)&=& E_0 + 2 \Omega - \Delta    \\
   E(70,1-) &=& E_0 + \Omega    \\
   E(56,0+) &=& E_0    
\end{eqnarray}
where the standard Coulomb plus linear potential gives $\Delta >0$, and
indeed gives $\Delta$ comparable to $\Omega$.  Here $E_0$, $\Omega$,
and $\Delta$ are three constants which are rigorously related by theory to 
the usual three parameters of the interquark
potential:  an overall constant, $\alpha_s$, and the string tension $b$.  Since the
$[56,0^+]$,
$[70,1^-]$,  and $[56,2^+]$ supermultiplet states are very well known, $E_0$, $\Omega$,
and $\Delta$ are known, and one can confirm {\it experimentally} that $\Delta$ is
comparable to
$\Omega$, i.e., that the $[56',0^+]$ supermultiplet should be close to the $[70,1^-]$.

     The second effect is OGE. In the Isgur-Karl model, spin-spin interactions from OGE
dominate the departures of the
$S=0$ masses from the $SU(6)$ limit and have a very simple structure.  Using the
zeroth order harmonic oscillator wavefunctions (I have once again arranged these equations
so that the lightest state is at the bottom of the array):,
\begin{eqnarray}
  \Delta'            &=&  E_0  +\Omega -\Delta + 5/8 d   \\
   N'                 &=&  E_0  +\Omega -\Delta - 5/8 d     \\  
  ^4N^* = ^2\Delta^*    &=&  E_0 +\Omega + 1/4 d     \\
  ^2N^*                &=&  E_0 +\Omega - 1/4 d     \\
   \Delta             &=&  E_0 + 1/2 d     \\
   N                 &=&  E_0 - 1/2 d ~~.        
\end{eqnarray}
Thus since from $\Delta-N$ we know $d=300$ MeV, with $\Omega$ approximately equal to
$\Delta$, one expects the $N^*(1440)$
to be roughly $3/8 d$ or about 110 MeV below the $N^*(1535)$ as observed.

     While a very satisfactory starting point, this simple story certainly 
has some well-known problems.  With $\Omega$ approximately equal to $\Delta$, the use of
perturbation theory in the anharmonicity is suspect and indeed realistic potentials all
have $E(56',0^+) > E(70,1^-)$.  That $\Delta' -N'$ is larger than $\Delta -N$ is also an
artifact of the harmonic limit.  Thus more complete analyses of the OGE model tend to
leave the $N^*(1440)$ too high in mass.  I will comment on this discrepancy below, but for
now note that these two states are naturally very close to each other, and are certainly
{\it not} expected to be split by a harmonic-like gap of $\Omega$.  

     Now consider the $S=-1$ sector, which is used as the key to defining the
``level ordering problem" of the Isgur-Karl model.  The OPE papers \cite{GR} note that in
this case the lightest states of the $N=$ 0, 1, and 2 bands, namely the $\Lambda$,
$\Lambda^*(1405)$, and
$\Lambda^*(1600)$, have the ``normal" level ordering  and claim that 
this reversal is inconsistent with the OGE picture.  The argument given is incorrect.  In
addition to the SU(3) partners $^2\Lambda_8$ and $^4\Lambda_8$ of the octet  $N^*$ states
$^2N$ and $^4N$, the $N=1$ band contains a new SU(3) singlet state $^2\Lambda_1$ {\it
that has no $N^*$ counterpart}.  The OGE spin-spin interaction in the SU(3) limit predicts
that the
$^4\Lambda_8$,
$^2\Lambda_8$, and $^2\Lambda_1$ states will have hyperfine interactions of $+1/4 d$,
$-1/4 d$, and $-3/4 d$, respectively: the new singlet state $^2\Lambda_1$ is the {\it
most highly stabilized state in the entire negative parity spectrum.} It automatically
drops substantially relative to the octet states, and thus OGE naturally gives a low-lying
negative parity state.  

     Thus the claim that the Isgur-Karl model (with its OGE) cannot explain the
observed level orderings, and especially the reversal of the $\Lambda$'s relative to the
$N^*$'s, is incorrect.  A more productive discussion might focus on the accuracy of 
the predicted spectrum, since neither the $N^*(1440)$ nor the $\Lambda^*(1405)$ mass
is very accurately predicted. As we have seen, a lot of physics is
missing from the valence quark model. In particular, I believe that one should not expect
flux tube model spectroscopy to work to better than 50 MeV until we have learned how to
treat the couplings to continua correctly \cite{NIonThresholds}.

%%%%%%%%%%%%%%%%%%%%%%%%%%%%%%%%%%%%%%%%%
\section{Conclusions}
%%%%%%%%%%%%%%%%%%%%%%%%%%%%%%%%%%%%%%%%%

   The study of $N^*$'s can provide us with critical insights into the nature of QCD in the
confinement domain. It is becoming clear that, in this domain, constituent quarks and flux
tubes are the appropriate low-energy degrees of freedom. However, there
are still many very important questions to be resolved on the nature of the short-distance
interactions of the quarks.  I expect that $N^*$'s, as has always been  the
case historically,  will in the future play a pivotal role in answering these questions.

%%%%%%%%%%%%%%%%%%%%%%%%%%%%%%%%%%%%%%%%%
\section{Acknowledgements}
%%%%%%%%%%%%%%%%%%%%%%%%%%%%%%%%%%%%%%%%%

     I am very
grateful to Volker Burkert for holding open until the very last minute an opportunity for
me to speak at $N^*$2000, despite a sudden illness which prevented me from participating
as planned, and to the members of the scheduled round table discussion for the graceful way
they accomodated themselves to these last minute changes.

    This work was supported by DOE contract DE-AC05-84ER40150 under which the
Southeastern Universities Research Association (SURA) operates the Thomas
Jefferson National Accelerator Facility.

\end{document}